\documentclass[%
 reprint,
superscriptaddress,
 amsmath,amssymb,
 aps,
prl,
floatfix,
]{revtex4-2}

\usepackage{graphicx}
\usepackage{dcolumn}
\usepackage{bm}

\usepackage{lipsum}
\usepackage{xcolor}

\begin{document}

\preprint{APS/123-QED}

\title{Arbitrary vector beam generation in semiconductor quantum dots}

\author{Samit Kumar Hazra}
\email{samit176121009@iitg.ac.in}
\affiliation{Department of Physics, Indian Institute of Technology Guwahati, Guwahati 781039, Assam, India}
\author{P. K. Pathak}
\email{ppathak@iitmandi.ac.in}
\affiliation{School of Physical Sciences, Indian Institute of Technology Mandi, Mandi 175005, Himachal Pradesh, India}
\author{Tarak Nath Dey}
\email{tarak.dey@iitg.ac.in}
\affiliation{Department of Physics, Indian Institute of Technology Guwahati, Guwahati 781039, Assam, India}

\date{\today}

\begin{abstract}

We have proposed an arbitrary vector beam (VB) generation scheme in a thin disk-shaped quantum dot (QD) medium considering phonon interaction. The QD biexciton system exhibits interplay between first and third-order nonlinear susceptibility between two orthogonal circular polarisation transitions. Three QD transitions are coupled with one applied weak and two strong control orbital angular momentum (OAM) carrying fields. Therefore, the applied field experiences absorption, and a new field with the desired OAM is generated via four-wave mixing (FWM). These two orthogonal field superpositions produce VB at the QD medium end. We have also demonstrated the polarization rotation of a VB by changing only the relative control field phase. Additionally, we have analyzed the effect of temperature on the VB generation.

\end{abstract}

\maketitle


\section{INTRODUCTION}
\label{sec:intro}

The spatial polarization inhomogeneity of vector beam(VB) light has gained research interest in the optics community due to its fundamental applications, including high-density optical communication and super-resolution imaging. Such VB generation requires vector superposition of two orbital angular momentum-carrying Laguerre-Gaussian(LG) modes with orthogonal polarization \cite{Zhan:09, Beckley:10, Galvez:12}. The solution of the paraxial wave equation provides LG modes with an optical vortex structure and, therefore, capable of carrying the orbital angular momentum (OAM) \cite{PhysRevA.56.4064}. This class of scalar beams with homogeneous polarisation distribution has been studied extensively in the literature. However, the VB is a relatively new and unexplored concept. Based on the OAM of each component, VBs are classified into two groups: full Poincare (FP) beams \cite{Beckley:10}, and cylindrical vector (CV) beams \cite{Zhan:09}. The FP and CV beams consist of components with one nonzero OAM and two equal and opposite OAM, including examples such as lemon, star, web, radial, azimuthal, and spiral VB, respectively \cite{Senthilkumaran2020}.\\

The CV beams have wide applications in the various fields of science and technology. The unique property of the CV beam demonstrates high numerical aperture (NA) focusing \cite{Youngworth:00, Zhan:02}, resulting in a significantly small spot size and beating the theoretical focusing limit for scalar beams \cite{PhysRevLett.91.233901}. Several other applications have also been reported, including optical trapping \cite{Michihata2009, Kozawa2010, Roxworthy2010}, super-resolution microscopy \cite{Kozawa2018, Török2004}, optical communication \cite{Liu2018, Willner2018}, and high harmonic generation \cite{Biss2003, Bautista2012, Bautista2013}. Subsequently, the theoretical prediction suggested a robust VB propagation through atmospheric turbulence \cite{Cheng2009, Gu2009, Cox2016, Wei2015}. The nonseparability of VB can be used to encode information for optical communication \cite{Milione2015}. Further, VBs provide an infinite-dimensional Hilbert space platform associated with OAM to study quantum entanglement \cite{PhysRevA.94.030304}, quantum key distribution (QKD) protocols \cite{PhysRevA.93.032320}, and  quantum cryptography \cite{Sit2017} for high-security communication. 

Due to the growing popularity of VB and its application in various fields, the VB generation technique has also gained much attention. Conventionally, the VB generation requires an interferometer setup with precise alignment between two components \cite{Maurer2007, Kalita2016}. Some initial attempts on radially polarised beam generation methods rely on an image-rotating resonator \cite{Armstrong2003} and double interferometer \cite{Tidwell1993}. Several other systems consider the Sagnac-like interferometer \cite{Wang2016}, Twyman-Green method \cite{Fu2015}, and Wollaston prism \cite{Xin2012}. In addition, several other platforms, like polarization gratings \cite{Sakamoto2017}, optical ﬁbers \cite{Viswanathan2009}, and ring resonators \cite{Schulz2013} have been used for vector vortex beam generation. The current commercial production of VB utilizes advanced optical elements such as a spatial light modulator (SLM) \cite{Rosales-Guzmán2017}, digital micromirror devices (DMDs) \cite{Gong2014}, and liquid crystal-based q-plates \cite{PhysRevA.100.053812}. 

Although many different VB generation techniques are available, almost none match the current quantum architecture or satellite communication requirement due to their large setup and high power consumption. The recent development of nanotechnology has led to various systems for vortex beam generation, such as integrated silicon-chip-based vortex beam emitters \cite{qiu2017vortex}, vortex vertical-cavity surface-emitting lasers (VCSELs) \cite{Toda2017}, angular gratings \cite{cai2012integrated}, micro-nano-OAM laser emitters \cite{miao2016orbital}, and various metasurface designs \cite{Wang2018}. Other microsystems regarding vortex beams have been invented, such as vector vortex on-chip generators \cite{Sun2016} and parallel OAM processors \cite{Chen2018}. These vortex beams could be helpful for VB generation in a microstructure.

This paper explores the possibility of arbitrary VB generation in a thin disk-shaped semiconductor quantum dot (QD) medium. In contrast, the QDs show full potential for this scheme because of the predetermined fabrication technology, tiny footprint, and ultra-low power consumption. Though the QD shows some remarkable advantages, lattice vibration is inevitable due to its sloid-state nature under the environment temperature. This temperature-dependent lattice vibration leads to the longitudinal acoustic phonon interaction with deformation potential. The phonon interaction with the QD exciton state results in various distinct features, such as dephasing \cite{PhysRevB.60.R2157}, zero phonon line broadening \cite{PhysRevB.70.201301}, off-resonant cavity feeding \cite{PhysRevB.83.165313}, and Mollow triplet \cite{PhysRevLett.106.247403} observed. To study the system dynamics, polaron transformation and corresponding master equation have been considered for the phonon interaction \cite{POLARON1, POLARON2, POLARON3}. In the literature, Hsu et al. \cite{PhysRevA.83.053819} reported the controllable propagation of an optical field due to the cross-talk between first and third-order nonlinear susceptibility in a four-level diamond-like system. Later, a similar QD system shows the  OAM transfer from the control field to the generated field via four-wave mixing (FWM) \cite{PhysRevA.101.023821}. Motivated by these works, we can find a scheme where one part of the field gets absorbed by two-level absorption, and the other part experiences gain due to ladder transparency associated with the first, and third-order nonlinear susceptibility, respectively. In this scenario, two orthogonal polarisation components with the desired OAM have been generated, and their superposition leads to the arbitrary VB generation.

\section{MODEL SYSTEM}
\label{sec:model}
In this paper, we present a simple and practical scheme for vector beam generation in a quantum dot(QD) medium. The medium comprises a few self-organized InGaAs QD layers separated by GaAs wetting layers created by the Molecular Beam Epitaxy (MBE) \cite{Sugaya2012}. In the self-organized process, all the QDs are not identical in size. Thus, different QD have different emission frequencies, which causes inhomogeneous broadening in the system. However, we have assumed all the QDs to be identical throughout the paper to simplify our problem. Therefore, a single QD up to two excitons exhibits a four-level diamond-like energy level system. The system consists of a ground-state $\vert g \rangle$, two exciton states $\vert x \rangle$, $\vert y \rangle$ and a biexciton state $\vert u \rangle$ with corresponding energy $\hbar \omega_{i}$ where $i \in \{g,x,y,u\}$. Normally, two exciton states exhibit unequal energy due to the underlying asymmetry of  QD. The frequency difference between two exciton states is known as fine structure splitting(FSS), defined by $\delta_{x} = \omega_{x} - \omega_{y}$. The biexciton state is a bound state of two exciton states. The energy required to form a biexciton state is known as biexciton binding energy, expressed as $\Delta_{xx} = \omega_{x} + \omega_{y} - \omega_{u}$. Various methods can reduce the FSS, such as thermal annealing and application of external electric and magnetic fields \cite{Plumhof2012}. However, the uniaxial stress applied by the piezoelectric process is the most convenient technique for making it zero \cite{PhysRevLett.106.227401}. Our system considers the FSS zero to access the circular polarization basis of the QD transition to produce vector beam components. We also choose very high biexciton binding energy to differentiate the frequency between probe and control beams required for detection.

\begin{figure}[h]
   \includegraphics[scale=0.33]{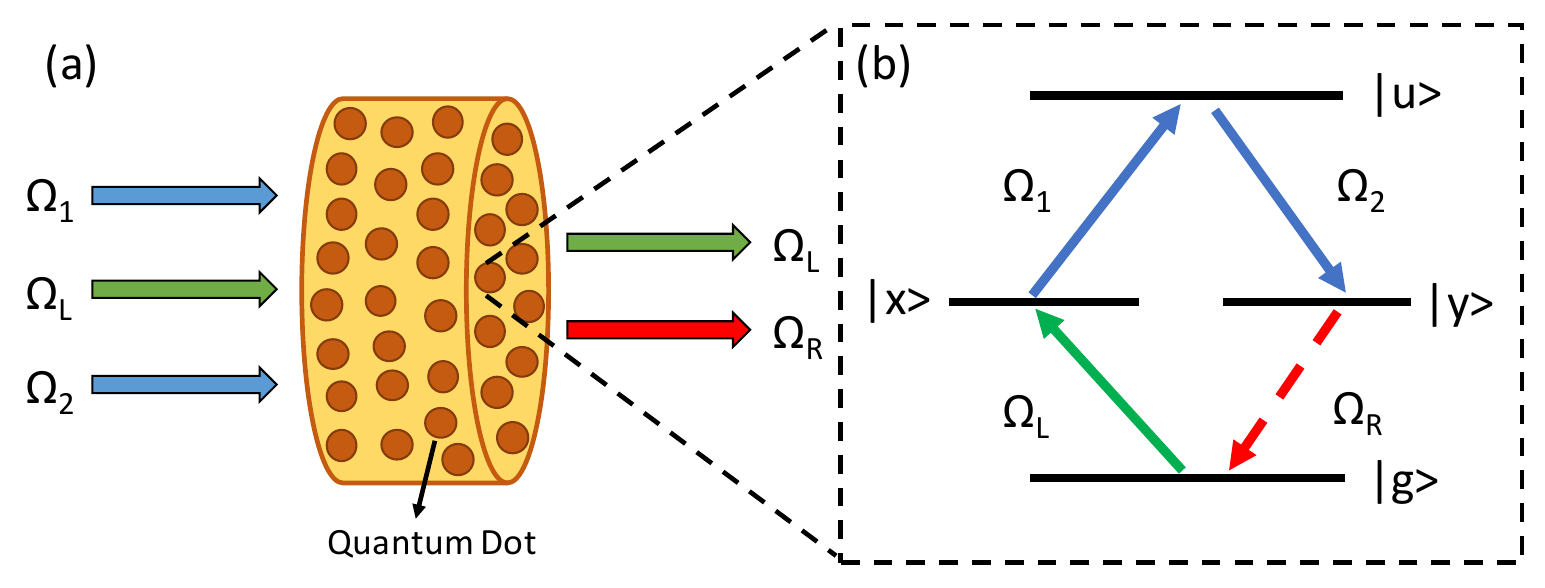}
    \caption{a) The schematic diagram of one weak probe field $\Omega_{L}$ and two strong control fields $\Omega_{1},\Omega_{2}$ passing through a thin disc shape QD medium and generating an FWM field $\Omega_{R}$. On the right-hand side, $\Omega_{L}$ represented by the green arrow is the transmitted probe field, and the two strong control field's presence is not shown in the figure due to far-detuned frequencies.  b) The diamond-shaped energy level structure of QD interacting with the applied fields with corresponding Rabi frequencies.}
    \label{Fig.1}
\end{figure}
 
In this scheme, one weak probe field $\vec{E}_{L}$ and two strong control fields $\vec{E}_{1}, \vec{E}_{2}$ propagate through a thin disc-shaped structure made of QD layers and generate a weak field $\vec{E}_{R}$ via four-wave mixing(FWM) process, as shown in Fig.1(a). The corresponding electric fields are  given by
\begin{equation}\label{eq:field}
\vec{E}_{j}(\vec{r},t)=\hat e_{j}\mathcal E_{0j}(\vec{r})e^{i(k_j z-\omega_j t)}+c.c., \hspace{0.2cm}	j\in \{L,R,1,2\}
\end{equation}
where $\mathcal E_{0j}(\vec{r})$ is the transverse variation of envelope, $k_{j}=\omega_{j}/c$ is the propagation constant, $\omega_{j}$ is the frequency and $\hat e_{j}$ is the polarisation vector of the quasi-monochromatic field. It is worth mentioning that the optical field $j=L,2$ and $j=R,1$ have left and right circular polarisation matching with the corresponding QD transition. Therefore, the Rabi frequencies after both rotating wave and dipole approximation have the form  $\Omega_{j}=-\vec{d}.\hat e_{j}\mathcal E_{0j}/\hbar $ where $j \in \{L, R,1,2\}$ and $\vec{d}$ is the dipole moment vector of the QD transition.

Fig.1(b) shows the schematic energy levels of the QD interacting with the various optical fields. Initially, all the population is in the ground state $\vert g \rangle$.
Then, we apply a weak probe field $\Omega_{L}$ and two strong control field $\Omega_{1}$, $\Omega_{2}$ to the $|g\rangle \rightarrow |x\rangle$, $|x\rangle \rightarrow |u\rangle$ and $|u\rangle \rightarrow |y\rangle$ transition, resulting in a small population redistribution from $|g\rangle \rightarrow |x\rangle \rightarrow |u\rangle \rightarrow |y\rangle$. Finally the weak four-wave mixing field $\Omega_{R}$ will be generated through $|y\rangle \rightarrow |g\rangle$ transition obeying the phase matching condition $k_{R} = k_{L} +k_{1} - k_{2}$. In this process, the population returns to the ground state, and the generated field $\Omega_{R}$ carries the transferred OAM from the control beam.
The non-zero intensity of $\Omega_{L}$ and $\Omega_{R}$ are required at the medium's output end to generate a vector beam. Therefore, we have considered a thin QD medium, which helps the weak probe $\Omega_{L}$ to reach the output end before being entirely absorbed by the medium. Hence, a fraction of the $\Omega_{L}$ is transmitted through the medium, and the coherently absorbed part is converted to $\Omega_{R}$ having OAM, resulting in a vector beam generation. Any semiconductor QD medium is inevitable to the surrounding lattice vibration caused by environmental temperature. Therefore, the QD system dynamics get modified due to the quantized mode of thermal vibration, i.e., acoustic phonon. The phonon bath consists of a collection of infinite closely spaced harmonic oscillators. Thus, annihilation
and creation operators of the kth phonon mode with frequency $\omega_{k}$ will be $b_{k}$ and $b_{k}^{\dagger}$. We consider the QD-phonon interaction in the total Hamiltonian to solve the QD dynamics in more detail.

As we have considered zero FSS in our system, the frequency of both the exciton will be equal $\omega_{x}=\omega_{y}$. Therefore, the applied field frequencies have the following structure $\omega_{L}=\omega_{R}=\omega_{p}$ and $\omega_{1}=\omega_{2}=\omega_{c}$ to remove the explicit time dependency from the interaction  Hamiltonian. The interaction picture Hamiltonian with phonon interaction in a suitable unitary transformation frame is
\begin{eqnarray}
H &&= -\hbar\delta_{p}(\sigma_{xx} + \sigma_{yy}) - \hbar(\delta_{p}+\delta_{c})\sigma_{uu}\nonumber\\ 
&&+\hbar\left( \Omega_{L}\sigma_{xg} +\Omega_{1} \sigma_{ux}  + \Omega_{2}\sigma_{uy} +\Omega_{R}\sigma_{yg} + H.c. \right) \\~\nonumber 
&& + \hbar\sum_{k}\omega_{k}b_{k}^{\dagger}b_{k}+\sum_{i=x,y,u}\lambda_{k}\sigma_{ii}(b_k+b_k^{\dag})\label{uhu},
\end{eqnarray}
where $\delta_{p} = \omega_{p} - \omega_{x}$,
$\delta_{c} = \omega_{c}-(\omega_{u}-\omega_{x})$  are  the detunigs of corresponding QD transition and $\lambda_{k}$ is the coupling streangth. The QD projection operators defined by $\sigma_{ij}=\vert i\rangle \langle j\vert$ where $\vert i\rangle$ and $\vert j \rangle$ are the QD states. To deal with the entire order of phonon interaction, we choose the polaron transformation, $H^{\prime}=e^P H e^{-P}$ where
$P=\sum_{i=x,y,u}\sigma_{ii}\sum_{k}\lambda_{k}\left(b_k^{\dag} - b_{k}\right)/\omega_{k}$. The transformed Hamiltonian decouple the system Hamiltonian $H_{S}$ from the bath Hamiltonian $H_{B}$ and the QD-bath interaction Hamiltonian $H_{I}$ with renormalized Rabi frequency given by $H^{\prime}=H_{S}+H_{B}+H_{I}$ where
\begin{eqnarray}
H_{S} &&= -\hbar\Delta_{p}(\sigma_{xx} + \sigma_{yy}) - \hbar(\Delta_{p}+\Delta_{c})\sigma_{uu} +\langle B\rangle X_{g},\\
H_{B} &&=\hbar\sum_k\omega_k b_k^{\dag}b_k,\\
 H_{I} &&=\xi_gX_g+\xi_uX_u.
\end{eqnarray}
After this transformation, previous detunings are redefined into effective detunings $\Delta_{p}$ and $\Delta_{c}$ by summing up the additional polaron shift $\sum_k\lambda_{k}^2/\omega_k$. The polaron frame system operators are 
\begin{eqnarray}
X_{g}&&=\hbar\left( \Omega_{L}\sigma_{xg} + \Omega_{1}\sigma_{ux}+\Omega_{2}\sigma_{uy} + \Omega_{R}\sigma_{yg} \right) + H.c.\\
X_{u}&&=i\hbar\left( \Omega_{L}\sigma_{xg} + \Omega_{1}\sigma_{ux}+\Omega_{2}\sigma_{uy} + \Omega_{R}\sigma_{yg} \right) + H.c. ,
\end{eqnarray}
and the phonon bath fluctuation operators are
\begin{equation}
 \xi_{g} = \frac{1}{2}\left( B_{+} + B_{-} -2\langle B\rangle  \right)\quad
 \xi_{u} = \frac{1}{2i}\left( B_{+} - B_{-} \right),
 \end{equation}
where $B_{+}$,$B_{-}$ are the phonon  displacement operator. 
The expression for phonon displacement operators in terms of phonon creation and annihilation operators is $B_{\pm} = $ exp$[\pm\sum_{k} \frac{\lambda_{k}}{\omega_{k}}\left( b_{k} - b_{k}^{\dagger}\right)]$. Hence, the expectation value of this operator at a temperature T with a phonon spectral density $J(\omega)$ provides us $\langle B_{+}\rangle =\langle B_{-}\rangle = \langle B\rangle = \text{exp}\left[-\frac{1}{2}\int_0^{\infty}d\omega\frac{J(\omega)}{\omega^2}
\coth\left(\frac{\hbar\omega}{2K_bT}\right)\right]$, where $K_{b}$ is the Boltzman constant. We have considered the experimentally verified phonon spectral density function $J(\omega)= \alpha_{p}\omega^3\exp[-\omega^2/2\omega_b^2]$ in our calculation, where the parameters $\alpha_p$ and $\omega_b$ are the electron-phonon coupling and cutoff frequency, respectively.

Now, to derive the polaron Master Equation(ME), we model a small system $H_{S}$ placed in a large phonon reservoir $H_{B}$ and interacting with the reservoir $H_{I}$. We apply the Born-Markov approximation to the system density matrix equation, which considers up to the second order in exciton-photon coupling. The phonon reservoir is chosen to be in thermal equilibrium to factorize the density matrix in the initial time. The time convolutionless ME for the reduced density matrix of a QD-field system in the presence of a phonon environment is given by
\begin{align}
\dot{\rho} &=-\frac{i}{\hbar}[H_{S},\rho] -\sum_{i=x,y}\left(\frac{\gamma_1}{2}{\cal L}[\sigma_{gi}]
+\frac{\gamma_2}{2}{\cal L}[\sigma_{iu}]\right)\rho  ~\nonumber\\
& -\sum_{i=x,y,u}\frac{\gamma_d}{2}{\cal L}[\sigma_{ii}]\rho -\frac{1}{\hbar^2}\int_0^{\infty}d\tau\sum_{j=g,u} \times ~\nonumber\\
& \big( G_j(\tau)[X_j(t),e^{-iH_{S}\tau/\hbar}X_j(t)e^{iH_{S}\tau/\hbar}\rho(t)]+H.c. \big), \label{meq}
\end{align}
where $\gamma_1$, $\gamma_2$ are the spontaneous decay rates of exciton and biexciton states, $\gamma_d$ refers to pure dephasing rate, and $G_{g/u}$ corresponding to the polaron Green functions. The pure dephasing process occurs due to the imperfectness of the system and is responsible for the zero-phonon line broadening in QD, which also depends on temperatures. The Green functions can be calculated from the correlation between bath fluctuation operators at time $\tau$ as $G_g(\tau)=\langle B\rangle^2\{\cosh[\phi(\tau)]-1\}$, $G_u(\tau)=\langle B\rangle^2\sinh[\phi(\tau)]$ where phonon correlation $\phi(\tau)=\int_0^{\infty}d\omega\frac{J(\omega)}{\omega^2}\left[\coth\left(\frac{\hbar\omega}{2K_bT}\right)\cos(\omega\tau)-i\sin(\omega\tau)\right]$. The  Lindblad superoperator $\cal L$ in the ME has the well-known form ${\cal L}[{\cal O}]\rho ={\cal O}^{\dagger}{\cal O} \rho - 2 {\cal O} \rho {\cal O}^{\dagger} +\rho \cal O^{\dagger}\cal O $ where $\cal O$ represent any arbitrary operator.

According to the Maxwell equation, any electromagnetic field interacting with QD creates induced polarisation $\vec{P}(z,t)$ due to its dipole alignment. This induced polarisation is associated with the coherence term of density matrix elements $\rho_{xg}$ and $\rho_{yg}$ expressed as $\langle x\vert \rho \vert g\rangle $, $\langle y\vert \rho \vert g\rangle $ for two weak probe fields $\Omega_{L},\Omega_{R}$. The induced polarization amplitude is written as
\begin{align}
{\cal P}_{xg}(z,t) = N d \hspace{1pt} \rho_{xg},\\
{\cal P}_{yg}(z,t) = N d \hspace{1pt} \rho_{yg},
\end{align}
where $N$ is the QD volume number density. By applying slowly varying envelope approximation to the Maxwell wave equation and making  a frame transformation $\tau =  t - z/c, \zeta = z$, we get the propagation equation 
\begin{align}
\frac{\partial}{\partial \zeta}  \Omega_{L}(\zeta,\tau) = i\hspace{1pt}\eta \hspace{1pt} \rho_{xg},\\
\frac{\partial}{\partial \zeta}  \Omega_{R}(\zeta,\tau) = i\hspace{1pt}\eta \hspace{1pt} \rho_{yg}, \label{prop_eq}
\end{align}
where the coupling constant $\eta = - 3N\lambda^{2}\gamma_{1}/4\pi$ and $\lambda$ is the central wavelength of the QD transition. We solve the master equation (\ref{meq}) and propagation equation simultaneously to calculate the generated field numerically using Quantum Optics Toolbox\cite{QOTOOL} in MATLAB. Now, we define the transmitted and generated electric fields $\mathcal{E}_{L}$, $\mathcal{E}_{R}$ at the output end of the medium. The generated vector beam (VB) in a cylindrical coordinate system is 
\begin{equation}
\vec{E}(r,\phi,z) = \mathcal{E}_{L}(r,\phi,z)\hat{e}_{L}+ \mathcal{E}_{R}(r,\phi,z)\hat{e}_{R},\label{vbcont}
\end{equation}
where $\mathcal{E}_{L}(r,\phi,z) = \cos(\alpha)LG_{0}^{l_{L}}, \ \textnormal{and}\ \mathcal{E}_{R}(r,\phi,z) = e^{i\theta}\sin(\alpha)LG_{0}^{l_{R}}$ are the Laguerre- Gaussian modes of left and right circular polarization with two controlling parameters, $\alpha$ and $\theta$ known as relative amplitude and phase. The considered Laguerre- Gaussian modes with zero radial index $LG_{0}^{l_{i}}(i \in L,R)$ has the form 
\begin{align}
LG_{0}^{l_{i}}(r,\phi,z) =E^{0}_{i} \sqrt{\frac{2}{\pi\vert l_{i}\vert!}} \left(\frac{r\sqrt{2}}{w(z)}\right)^{\left|l\right|} \times ~\nonumber\\
e^{-\frac{r^{2}}{w^{2}(z)}}e^{\frac{ik_{i}^{f}n_{i}r^{2}z}{2\left( z^{2} + n_{i}^{2}z_{R}^{2} \right)}} e^{il_{i}\phi-i(|l|+1)\eta(z)+ik_{i}^{f}n_{i}z}.
\end{align}
Used notations are the maximum field amplitude $E^{0}_{i}$, angular momentum $l_{i}$, azimuthal angle $\phi$, Rayleigh length $z_{R} = k_{i}^{f}w_{0}^{2}/2$, beam waist $w_{0}$, wave number $k_{i}^{f}$, beam waist at $z$ distance $w(z)= w_{0}\sqrt{1 + z^{2}/n_{i}^{2}z_{R}^{2}}$,  refractive index $n_{i}$, Gouy phase $(|l|+1)\eta(z)$ where $\eta(z)=\tan^{-1}(z/n_{i}z_{R})$. We adopted the Stokes parameter formalism in the circular polarization basis to visualize the polarization distribution in a transverse plane given by 
\begin{align}
S_{0} &= \vert \mathcal{E}_{L}\vert^{2} + \vert \mathcal{E}_{R}\vert^{2},\hspace{1mm} S_{1} = 2 \text{Re}[\mathcal{E}_{L}^{*}\mathcal{E}_{R}], ~\nonumber\\
S_{2} &= 2 \text{Im}[\mathcal{E}_{L}^{*}\mathcal{E}_{R}],  \hspace{4.5mm} S_{3} = \vert \mathcal{E}_{L}\vert^{2} - \vert \mathcal{E}_{R}\vert^{2}. \label{stoke}
\end{align}
With the help of Stokes parameters, the ellipticity $\chi$ and orientation $\psi$ of the degree of polarization at any given point can be calculated as follows
\begin{equation}
\frac{S_{1}}{S_{0}} = \cos(2\chi)\cos(2\psi), \frac{S_{2}}{S_{0}} = \cos(2\chi)\sin(2\psi),\frac{S_{3}}{S_{0}} = \sin(2\chi).\label{elptic}
\end{equation}
We can derive the expression for ellipticity $\chi$ and orientation $\psi$ in terms of the stokes parameters by using Eq.{$\ref{stoke}$} and Eq.{$\ref{elptic}$} and are given by

\begin{eqnarray}
\chi &= \frac{1}{2}\sin^{-1}\left( \frac{S_{3}}{S_{0}}\right) = \frac{1}{2}\sin^{-1}\left( \frac{\vert \mathcal{E}_{L}\vert^{2} - \vert \mathcal{E}_{R}\vert^{2}}{\vert \mathcal{E}_{L}\vert^{2} + \vert \mathcal{E}_{R}\vert^{2}}\right),\\
\psi &= \frac{1}{2}\tan^{-1}\left( \frac{S_{2}}{S_{1}}\right) = \frac{1}{2}\tan^{-1}\left( \frac{Im[\mathcal{E}_{L}^{*}\mathcal{E}_{R}]}{Re[\mathcal{E}_{L}^{*}\mathcal{E}_{R}]}\right).
\end{eqnarray}

\section{RESULTS}

\begin{figure*}
   \includegraphics[width = 0.9\textwidth]{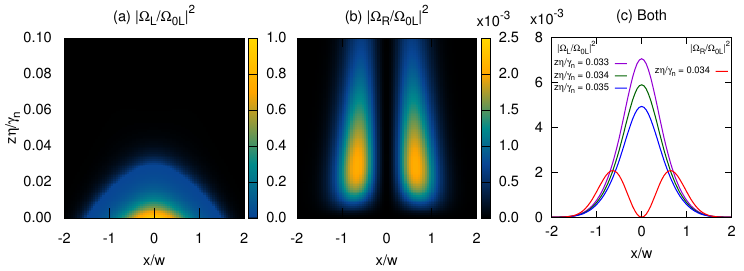}
    \caption{The intensity distribution of the applied left circular and generated right circular polarised field along the $x$-axis and its variation with the propagation distance $z\eta/\gamma_{n}$. a) Applied field intensity variation as a function of propagation distance for an LG beam with OAM $l_{L}$ =0. b) Generated FWM field intensity variation corresponding to the propagation length with a signature of OAM $l_{R}$ = 1. c) The relative intensity comparison between the two polarisation components for some specific $z\eta/\gamma_{n}$ satisfying the lemon VB condition. The generated intensity shows identical behavior for all three different propagation lengths; therefore, it is presented for only one propagation length. The considered applied field  Rabi frequencies are $|\Omega_{0L}| = 0.005\gamma_{n}$, $|\Omega_{01}| = 0.01\gamma_{n}$, $|\Omega_{02}| = 0.05\gamma_{n}$ resonantly coupled with the desired transition $\Delta_{p} = \Delta_{c} = 0$. The suitable beam waist of the three fields is $w_{L} = 1.0w$ and $w_{1} = w_{2} = 1.7w$ at a phonon bath temperature T = 5K. The relative phases are $\theta_{1}=\theta_{2}=0$.}
    \label{Fig.2}
\end{figure*}

In order to study the VB generation in a QD medium, we have considered some experimental parameters that are compatible with our model system. The QD structure possess volume number density $N = 1.5 \times 10^{19} m^{-3}$. The phonon bath temperatures T = 0, 5, 10, 20 K gives $\langle B \rangle$ = 1.0, 0.90, 0.84, 0.73 for phonon spectral distribution parameters $\alpha_{p}\gamma_{n}^{2}= 1.42\times 10^{-3}$ and $\omega_{b} = 10 \gamma_{n}$ with normalization frequency $\gamma_{n} = 100 \, \mu eV$. The relaxation and dephasing rates of the QD are taken to be $\gamma_{1} =\gamma_{2}=\gamma_{d}=\gamma = 0.01\gamma_{n}$. The scheme comprises one weak applied field $\Omega_{L}$ and two strong control fields $\Omega_{1}, \Omega_{2}$ with the general LG beam structure. According to our definition, the Rabi frequencies associated with the following fields can be written as 
\begin{align}
\Omega_{L} (r,\phi,z=0) &= \Omega_{0L} \left( \frac{r\sqrt{2}}{w_{L}}\right)^{\vert l_{L}\vert} e^{-\frac{r^{2}}{w_{L}^{2}}} e^{il_{L}\phi},\\
\Omega_{1} (r,\phi,z=0) &= \Omega_{01} \left( \frac{r\sqrt{2}}{w_{1}}\right)^{\vert l_{1}\vert} e^{-\frac{r^{2}}{w_{1}^{2}}} e^{i(l_{1}\phi+\theta_{1})},\\
\Omega_{2} (r,\phi,z=0) &= \Omega_{02} \left( \frac{r\sqrt{2}}{w_{2}}\right)^{\vert l_{2}\vert} e^{-\frac{r^{2}}{w_{2}^{2}}} e^{i(l_{2}\phi+\theta_{2})},
\end{align}
where $r$, $\phi$ corresponds to radial distance, azimuthal angle in cylindrical coordinates along with the beam waist $w_{i}$ and OAM $l_{i}$ $i\in\{L, 1, 2\}$. In general, we incorporated the constant relative phase of the two control fields compared to the applied field, denoted by $\theta_{1}$ and $\theta_{2}$. Subsequently, all the beam waist in the definition of three LG beams have normalized with a common beam waist $w$ with the chosen value of 100\ $\mu\text{m}$.
In a similar atomic configuration, the effect of third-order nonlinearity has been studied extensively for field propagation dynamics, both theoretically \cite{PhysRevA.83.053819} and experimentally \cite{PhysRevLett.100.203001}. It has been predicted that the interplay between the first and third-order nonlinearity of the medium leads to a controlled field propagation. Motivated by this work, a recent study shows the structured beam generation and OAM transfer between fields via FWM process \cite{Mallick2020}. In this direction, the implementation of the spatially structured transparency and transfer of optical vortices via four-wave mixing in a QD nanostructure was reported recently \cite{PhysRevA.101.023821}. According to this study, the Rabi frequency of the generated FWM fields depends on the one applied field and two control fields as $\Omega_{R} \propto \Omega_{L}\Omega_{1}\Omega_{2}^{*}$. From this equation, it is clear that the OAM of the generated field will be $l_{R} = l_{L} + l_{1} -l_{2}$. Now, we can make our first attempt to generate a lemon VB, where the OAM has to be $l_{L} = 0$  and $l_{R} = 1$. Therefore to fullfill this requirement, we have chosen $l_{L} = 0,l_{1} = 1,l_{2} = 0$. Although the OAM of the left and right circular components of the VB are generated correctly, now we have to focus on the relative strength between the two components denoted by the angle $\alpha$.

In Fig.\ref{Fig.2}(a), we display the weak applied field intensity distribution variation during propagation inside the QD medium. The applied field with OAM $l_{L}$ = 0 shows a Gaussian distribution along the $x$-axis at $z$ = 0. Consideration of only the $x$-axis is justifiable because of the radial symmetry of the applied field. We notice that the field intensity rapidly decreases with increasing propagation distance and becomes zero for long distances. This behavior follows the well-known Beer's law of weak field absorption in a two-level medium. The presence of the two-level absorption term in the susceptibility explains this kind of feature at resonance conditions. As the medium acts as a strong absorber for the weak applied field $\Omega_{L}$, the length of the medium should be small enough to get a viable output intensity at the medium end. Therefore, the absorbed energy transfers a small amount of the QD population from the ground state $\vert g \rangle$ to the exciton state $\vert x \rangle$. The first control field $\Omega_{1}$ couples the transition between $\vert x\rangle \leftrightarrow \vert u\rangle$ with OAM $l_{1}$ = 1 exchange the population to the biexciton state $\vert u \rangle$. Subsequently, the second control field does not carry any OAM couple the transition $\vert u\rangle \leftrightarrow \vert y\rangle$ resulting in the redistribution of the population into $\vert y \rangle$ state. In this whole process, the third-order nonlinearity of the QD medium couples to the three fields and produces a new field via the FWM process with the following structure $\Omega_{R} \propto \Omega_{L}\Omega_{1}\Omega_{2}^{*}$. From this expression, we notice that the OAM of the first control field $l_{1}$ = 1 transferred to the generated field $\Omega_{R}$ as OAM  $l_{R} = 1$. Fig.\ref{Fig.2}(b) shows the variation of the intensity distribution normalized with the applied field peak intensity along the propagation length for the generated field. We notice that the generated field intensity becomes zero at $z$ = 0 as the production of $\Omega_{R}$ is not started. After the propagation of a certain distance inside the QD medium, the generation of the   $\Omega_{R}$ started and hence began to show the intensity. Most importantly, one can notice that the intensity distribution shows a doughnut shape along the $x$-axis. It is understandable from the fact that the transfer of the OAM  $l_{1}$ = 1 to the generated field creates such a structure. Even though we can generate a desired OAM carrying field from this system, the generated field's intensity is much weaker than the applied field because of the third-order nonlinearity. Looking at the propagation axis, we understand that the generated field intensity grows from zero to a maximum value and then gradually decreases to zero for larger distances due to the medium absorption.

\begin{figure*}
   \includegraphics[width=0.9\textwidth]{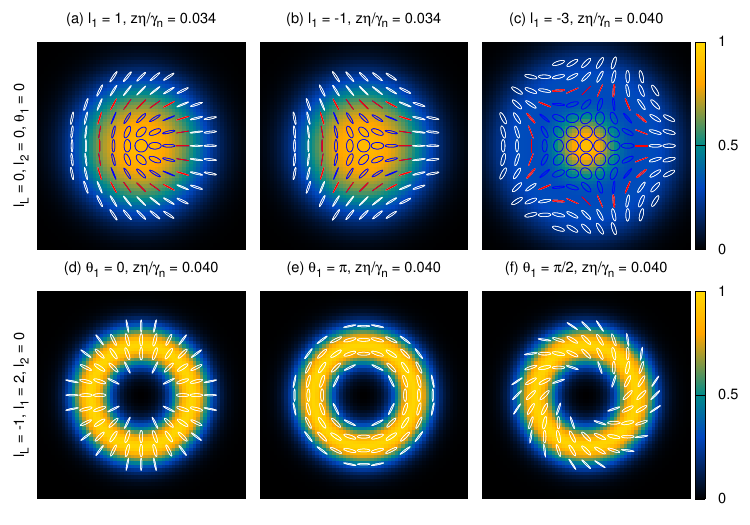}
    \caption{This figure illustrates the intensity and polarisation distribution for various generated VB in a transverse plane. The blue, white, and red color ellipse corresponds to the left and right circular and linear polarisation. All other parameters are the same as Fig.\ref{Fig.2} except the beam waist for (a)-(c) $w_{L} = 1.0w$ and $w_{1} = w_{2} = 1.7w$ and (d)-(f) $w_{L} = 0.8w$ and $w_{1} = w_{2} = 1.0w$.}
    \label{Fig.3}
\end{figure*}

We consider the medium length near $z\eta/\gamma_{n}$ = 0.03 to maximize the generated output intensity. Figure $\ref{Fig.2}$(c) depicts the variation of both the applied and generated field for a few specific propagation distances. It is worth mentioning that the relative intensity between the two components is an essential requirement in producing VB. For a lemon VB, the required peak intensity of the two components is roughly 3:1. We observed that the applied intensity decreases rapidly for three different propagation distances, but the generated field becomes constant. Among all the three distances, $z\eta/\gamma_{n}$ = 0.034 best fit with the actual lemon VB component distribution. Thus, our scheme can produce a lemon VB by applying three fields through a QD medium of width $z\eta/\gamma_{n}$ = 0.034.

From the previous study, it is clear that all the controlling parameters of the general VB in Eq.(\ref{vbcont}) are accessible in this scheme. We can choose any arbitrary OAM of $l_{L}$, $l_{1}$ and $l_{2}$ for the applied and control fields to generate a field carrying OAM $l_{R} = l_{L} +l_{1}-l_{2}$. Thus, any required OAM for the two components of a VB is achievable by considering a suitable combination of the three fields. The relative intensity between two VB components depends on $\alpha$
could be controlled by the propagation distance in our system. Further, the relative phase $\theta$  between two VB components can be regulated by changing the constant relative phase  $\theta_{1} - \theta_{2}$ between two control fields. Figure \ref{Fig.3} showcases the generated VB intensity and polarisation on a transverse plane with color plot and ellipse calculated from various stokes parameters. Therefore, each panel exhibits the distribution on the $xy$ plane for a specific propagation distance with different sets of parameters. Some of the most popular full Poincare VBs are presented in Fig.\ref{Fig.3}(a)-(c). For all the full Poincare VBs, we have only changed the OAM of the first control field by setting all other OAM and phase $l_{L}=l_{2}=0$, $\theta_{1}=\theta_{2}=0$. In Fig.\ref{Fig.3}(a), we observe a well-known lemon VB pattern for the first control field OAM $l_{1}$=1 and a propagation distance of $z\eta/\gamma_{n}$ = 0.034. This result is consistent with the previous analysis made on the lemon VB generation. The first panel of Fig.$\ref{Fig.3}$ shows flat-top Gaussian-like background intensity distribution because of the resultant intensity defined by the Stokes parameter $S_{0}$ of two VB components by looking at Fig.\ref{Fig.2}(c). The left circular polarisation components of the generated VB do not carry any OAM exhibit Gaussian distribution peaking at the center. In contrast, the generated right circular polarisation component is zero near the center because of OAM $l_{R}$ =1. Noticeably, the center of the VB shows a clear signature of the left circular polarisation denoted by blue ellipses. Further, going towards the edges, the Gaussian intensity decreases, but the doughnut intensity increases, resulting in linear polarisation and right circular polarisation. Overall, the inhomogeneous polarisation distribution looks like a lemon, and the total OAM is 1, named lemon VB under the FP VB category. In Fig.\ref{Fig.3}(b), we have considered everything similar to the previous panel except the first control field OAM $l_{1} = -1$, which transferred to the generated field OAM as $l_{R} = -1$ satisfy the star VB configuration. In star VB, the inhomogeneous polarisation distribution shows symmetry around three equally separated lines meeting at the center. The propagation distance is precisely the same as lemon vector beams because we have only changed the sign of the OAM, which does not affect the intensity distribution. In Fig.\ref{Fig.3}(c), we have displayed generated web VB by considering the control field OAM $l_{1}=-3$ with proper propagation distance $z\eta/\gamma_{n}$ = 0.040. This configuration differs entirely from the previous two, as we now consider the higher OAM to produce $l_{R}=-3$. For LG beams, the normalization factor contains $\sqrt{|l_{i}!|}$ on the  denominator is now effective for $|l_{i}|>1$. Therefore, the relative intensity between the web VB's left and right circular components has to obey an approximate ratio of 5:1 to produce the desired VB. The condition for the web VB is satisfied by taking a longer propagation distance in the QD medium compared to the others. The intensity distribution depicts a clear Gaussian peak at the center surrounded by the much lower-intensity doughnut distribution. Subsequently, the polarisation distribution shows a  web-like structure popularly known as web VB. Now, we move on to the other class of the VB, which has equal and opposite OAM-carrying components known as cylindrical VB. Fig.\ref{Fig.3}(d)-(f) presents three well-known cylindrical VBs generated from this scheme at a propagation distance $z\eta/\gamma_{n}$ = 0.040. Unlike the previous case, all the parameters are fixed here $l_{L}=-1, l_{1}=2, l_{2}=0, \theta_{2}=0$ except the first control field phase $\theta_{1}$. Therefore, the cylindrical VB can be produced by only changing the $\theta_{1}$. One important thing to notice here is that the applied field OAM $l_{L}=-1$ and the first control field OAM $l_{1}=2$ produces a resultant OAM $l_{R} = 1$ satisfy the condition for CVB. Figure \ref{Fig.3}(d)-(f) depicts the radial, azimuthal, spiral symmetry in polarization distribution on a transverse plane for $\theta_{1} = 0, \pi, \pi/2$, named radial azimuthal, spiral CVB. These CVBs have various applications in modern science and technology. For all three CVBs, we notice that the center dark spot is bigger than the conventional CVB. This behavior comes from the first control field OAM $l_{1}=2$ because the resultant field is the product of the $\Omega_{L}$ and $\Omega_{1}$, which left the signature of $l_{1}=2$ in the intensity distribution. Also, the eccentricity of the polarisation ellipse changes from higher to lower intensity region of the bright ring because of the nonidentical intensity distribution between the $\Omega_{L}$ and $\Omega_{R}$. Therefore, we could generate any arbitrary VB in this configuration by changing the controlling parameters.

\begin{figure}[h]
   \includegraphics[scale = 0.75]{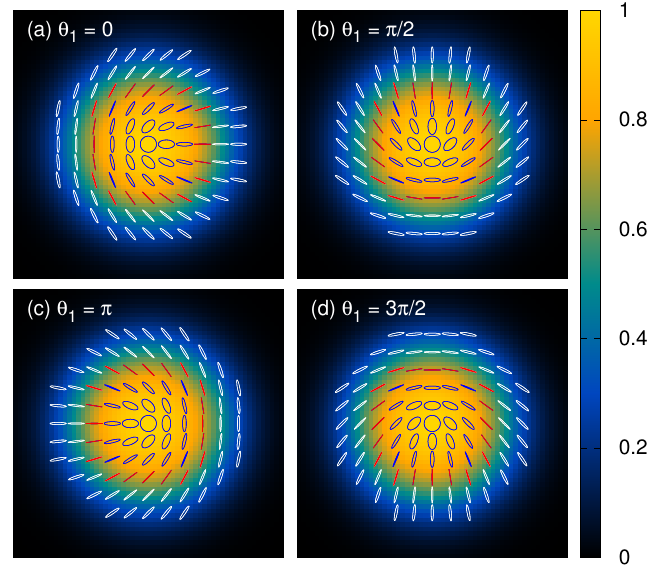}
    \caption{Controllable polarisation rotation of generated lemon VB by changing the first control field phase $\theta_{1}$. All other parameters are the same as Fig. \ref{Fig.2}.}
    \label{Fig.4}
\end{figure}

Figure \ref{Fig.4} displays the polarization distribution rotation of the generated lemon VB for four different values of the first control field phase angle $\theta_{1}$. For a lemon VB, the OAM of the left and right circular polarization components are $l_{L}=0$ and $l_{R}=1$. Therefore, the generated field creates a spatially dependent structured transparency in the QD medium's susceptibility, having one absorption and one gain peak. These peaks are identical and placed side by side with a common zero line passing through the center. The control field relative phases are defined by setting the reference point at the applied field $\Omega_{L}$ phase. Therefore, the second control phase $\theta_{2}=0$ indicates  the transfer of $\theta_{1}$ phase to the generated field $\Omega_{R}$. For Fig.\ref{Fig.4}(a), the relative phase between two components of the VB is zero, resulting in a lemon polarisation distribution pointing to the right side. By changing the phase $\theta_{1} = \pi/2$, the gain and absorption peak of the QD medium rotates ninety degrees clockwise in the transverse plane. This spatially dependent transparency rotation is responsible for the polarisation rotation of the lemon VB pointing upward, as shown in Fig.\ref{Fig.4}(b). Similarly, we can observe that the polarisation rotates accordingly for $\theta_{1} = \pi, 3\pi/2$.Therefore, the scheme is capable of any arbitrary polarisation rotation of a VB by tuning the first control field phase $\theta_{1}$.

\begin{figure}[h]
   \includegraphics[scale = 0.75]{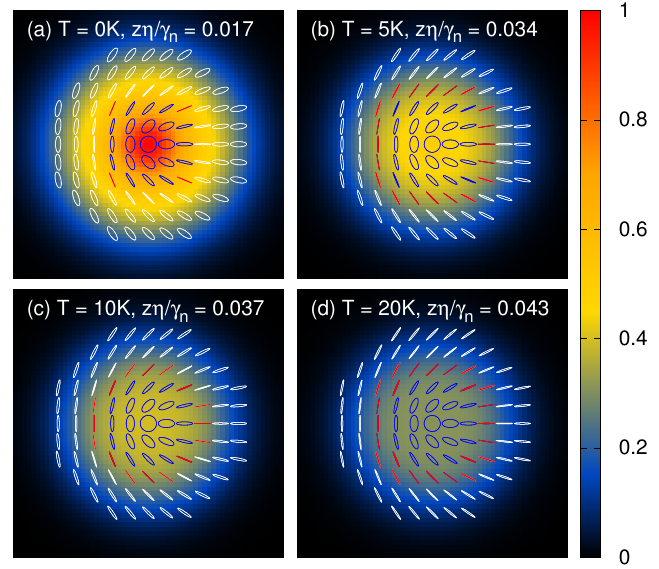}
    \caption{The figure illustrates the variation of transverse intensity and polarization distribution of a generated lemon VB for various phonon bath temperatures. All other parameters are the same as Fig.\ref{Fig.2}.}
    \label{Fig.5}
\end{figure}

Fig. \ref{Fig.5} displays the phonon bath temperature dependence on the VB generation. As mentioned earlier, the bath temperature of the system plays a vital role in the system dynamics. For increasing temperature, the renormalized Rabi frequency $\Omega_{i}\langle B\rangle$ gets reduced due to a smaller value of $\langle B\rangle$. Higher temperatures also introduce more dephasing in the system, which leads to the reduction of quantum coherence. As a result, phonon-induced decay rates get enhanced, resulting in low-intensity right circular polarisation component generation. Fig. \ref{Fig.5} (a) shows the lemon VB generation without phonon bath contribution in the Hamiltonian. We notice that the intensity of the generated beam is higher than all other temperatures. The ratio between the two components satisfies the lemon VB configuration for a small propagation distance. In the case of nonzero temperatures, the intensity of the generated VB diminished with increasing temperatures T = 5, 10, 20 K as depicted in Fig.\ref{Fig.5} (b)-(d). Accordingly, the system requires a longer propagation distance to satisfy the correct ratio between the VB components. From this analysis, it is clear that high temperatures reduce the output intensity of the generated VB and demand a long medium propagation length to produce it.

\section{CONCLUSIONS}
\label{conclud}

We have demonstrated a simple and compact system for generating arbitrary vector beams in a QD medium. This system incorporates the phonon interaction in the system Hamiltonian to provide a more realistic result. To deal with the phonon interaction, we make a polaron transformation on the total Hamiltonian and then find the master equation for the system density matrix. In this scheme, we can generate any arbitrary VB because of the access of the controlling parameters. We have shown that one can regulate the polarisation rotation of the generated VB  by changing the first control field phase. We have also explicitly studied the effect of the temperature on the VB generation. This scheme could potentially be applied to chip-based photonic circuits and free-space optical satellite communication.

\begin{acknowledgments}
T.N.D. gratefully acknowledges funding by Science and Engineering Research Board (Grant No. CRG/2023/001318). 
\end{acknowledgments}

\bibliography{paper3}

\end{document}